# CHORD GENERATION FROM SYMBOLIC MELODY USING BLSTM NETWORKS

**Hyungui Lim**[1,2], **Seungyeon Rhyu**[1] **and Kyogu Lee**[1,2]
[1]Music and Audio Research Group, Graduate School of Convergence Science and Technology
[2]Center for Super Intelligence
Seoul National University, Korea
`{goongding7, rsy1026, kglee}@snu.ac.kr`

## ABSTRACT

Generating a chord progression from a monophonic melody is a challenging problem because a chord progression requires a series of layered notes played simultaneously. This paper presents a novel method of generating chord sequences from a symbolic melody using bidirectional long short-term memory (BLSTM) networks trained on a lead sheet database. To this end, a group of feature vectors composed of 12 semitones is extracted from the notes in each bar of monophonic melodies. In order to ensure that the data shares uniform key and duration characteristics, the key and the time signatures of the vectors are normalized. The BLSTM networks then learn from the data to incorporate the temporal dependencies to produce a chord progression. Both quantitative and qualitative evaluations are conducted by comparing the proposed method with the conventional HMM and DNN-HMM based approaches. Proposed model achieves 23.8% and 11.4% performance increase from the other models, respectively. User studies further confirm that the chord sequences generated by the proposed method are preferred by listeners.

## 1. INTRODUCTION

Generating chords from melodies is an artistic process for musicians, which requires knowledge of chord progression and tonal harmony. While it plays an important role in music composition studies, the implementation of its process can be difficult especially for individuals who do not have prior experience or domain knowledge in musical studies. For this reason, the chord generation process often serves as an obstacle for novices who try to compose music based on a melody.

To overcome this limitation, automatic chord generation systems have been implemented based on machine learning methods [1, 2]. One of the most popular approaches for this task is probabilistic modelling, which commonly applies the hidden Markov model (HMM). A single-HMM is used with 12-semitone vectors of melody as observations and corresponding chords as hidden states [3, 4]. Allan and Williams trained a first-order HMM which learns from pieces composed by Bach, to generate chorale harmonies [5]. A more complex method is presented by Raczyński et al. [6], using time-varying tonalities and bigrams as observations with melody variables. In addition, a multi-level graphical model using tree structures and HMM is proposed by Paiement et al. [7]. Their model generates chord progressions based on the root note progression predicted from a melodic sequence. Forsyth and Bello [8] also introduced a MIDI based harmonic accompaniment system using a finite state transducer (FST).

Although the HMM has been successfully used for various tasks, it has several drawbacks. According to one of the assumptions of the Markov model, observations occur independently of their neighbors, depending only on the current state. Moreover, the current state of a Markov chain is only affected by its previous state. These drawbacks are also observable in chord generation from melody tasks because long-term dependencies exist in chord progressions and melodic sequences of Western tonal music [6].

Meanwhile, deep learning based approaches have recently shown great improvements in machine learning tasks of large datasets. Especially for temporal sequences, recurrent neural networks (RNN) and long short term memory (LSTM) networks have proven to be more powerful models than HMM in the field of handwriting recognition [9], speech recognition [10], and emotion recognition [11]. Nowadays, even music generation researches have increasingly adapted RNN/LSTM models in two major stream – one that aims to generate complete music sequences [12, 13], and the other which concentrates on generating music components such as melody, chord and drum sequence [14, 15]. We attempt an extended approach to the latter stream by implementing a chord generation system with a melody input.

In this paper, we implement a chord generation algorithm based on bidirectional LSTM (BLSTM) and evaluate its performance on reflecting temporal dependencies on melody/chord progressions by comparing with two HMM-based methods: a simple HMM, and deep neural networks-HMM (DNN-HMM). We then present the quantitative analysis and the accuracy results of the three models. We also describe the qualitative results based on subjective ratings provided by 25 non-musicians.

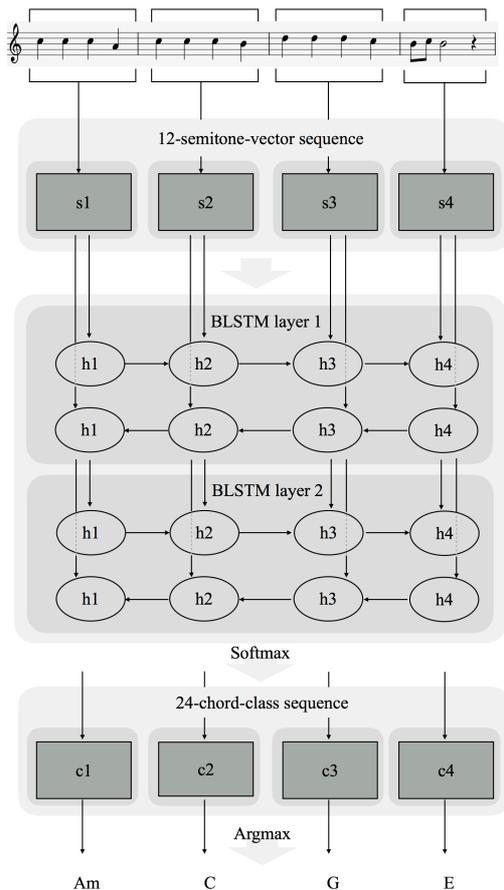

**Figure 1**. The overview of proposed system

The remainder of the paper is organized as follow. In Section 2, we explain the preprocessing step and the details of the machine learning methods we apply. Section 3 describes the experimental setup for evaluating the proposed approach. The experimental results are presented in Section 4, with additional discussions. Finally, we draw a conclusion followed by limitations and future works in Section 5.

## 2. METHODOLOGY

The method proposed in this paper can be divided into two main parts. The first part is a preprocessing procedure to extract input/output features from lead sheets. The other part consists of model training and a chord generation processes. We apply BLSTM networks for the proposed model and two types of HMM for the comparable models. The overall framework of our proposed method is shown in Figure 1.

### 2.1 Preprocessing

To extract appropriate features for this task, we first collect musical features such as time signature, measure (bar), key {fifths, mode}, chord {root, type} and note {root, octave, duration} from the lead sheets. These features are then represented in a matrix by concatenating rows, which respectively represent the musical features of a single note

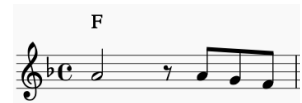

| time | measure | key_fifths | key_mode | chord_root | chord_type | note_root | note_octave | note_duration |
|------|---------|------------|----------|------------|------------|-----------|-------------|---------------|
| 4/4  | 1       | -1         | major    | F0         | major      | A0        | 4           | 8.0           |
| 4/4  | 1       | -1         | major    | F0         | major      | rest      | 0           | 2.0           |
| 4/4  | 1       | -1         | major    | F0         | major      | A0        | 4           | 2.0           |
| 4/4  | 1       | -1         | major    | F0         | major      | G0        | 4           | 2.0           |
| 4/4  | 1       | -1         | major    | F0         | major      | F0        | 4           | 2.0           |

**Figure 2**. An example of extracted data from a single bar.

as shown in Figure 2.

The generated data is then preprocessed in order to make an acceptable relation between melody input and chord output. All songs are in major key in the database and are transposed to C major key for data consistency. In other words, all roots of chords and notes are shifted to C major key to normalize different characteristics of melodies and chords in different songs.

Each song contains a time signature, which has a variety of meters such as 4/4, 3/4, 6/8, etc. The variety in time signature causes the imbalance of total note durations in a bar among different songs, so note durations are normalized by multiplying them with the reciprocal number of each time signature. After that, every note in a bar is stored into 12 semitone classes, without the octave information. Each class consists of a single value that accumulates the duration of the corresponding semitone in the bar.

Since the total number of chord types is quite large, if all of these chord types exist as independent classes, then each chord may not have enough samples. For such reason, all types of chords are mapped into one of two primary triads: major and minor. Each chord is represented with a binary 24-dimensional class to indicate the 24 major/minor chords.

### 2.2 BLSTM Networks

Recurrent neural networks (RNN) is a deep learning model, which learns complex networks not only by reconstructing the input features in a nonlinear process, but also by using the parameters of previous states in its hidden layer. A concept of "time step" exists in RNN, which is able to control the number of feedbacks on a recurrent process. This property enables the model to incorporate temporal dependencies by storing the past information in its internal memory, in contrast to a simple feedforward deep neural networks (DNN).

Despite such advantages of RNN models, there still exist problems regarding the long-term dependency. This is caused by vanishing gradient during the back propagation through time (BPTT) [16]. In the process of calculating the gradient of the loss function, the error between the estimated value and the actual value diminishes as the number of hidden layers increases. Thus, we instead use long short-term memory (LSTM) layers, which improve the limitation of storing long-term history with three multiplicative gates [17].

Generally, chords and melodies are formed in a sequential order, which is affected by both the previous and next order. Based on this, we can predict that if we reverse the lead sheet and train the musical progressions, a meaningful sequential context similar to the originals will appear. Hence, we apply a BLSTM so that the network can reflect musical context not only in forward but also in backward directions.

As shown in Figure 1, the input semitone vectors from each bar enter the network sequentially during the time step (i.e. a fixed number of bars) and emit the corresponding output chord classes in the same order. This is possible because the hidden layer in the network returns the output for each input. In order to train this sequence of multiple bars, we reconstruct our dataset by applying the window with the size of the time step and overlapping the window with the hop size of one bar. Each window, composed of multiple bars, is then used as a sample to train the network.

For our model, we build a time distributed input layer with 12 units, which represents the sequence of semitone vectors, 2 hidden layers with 128 BLSTM units, and a time distributed output layer with 24 units, which represents the sequence of chord classes. We empirically choose the number of hidden layers and units that yield the best result. We use hyperbolic tangent activation function for the hidden layers to reconstruct the features in a nonlinear process. We then apply the softmax function for the output layer to generate values corresponding to the probability of each class. Dropout is also employed with a rate of 0.2 on all hidden layers to prevent overfitting. We use mini-batch gradient descent with categorical cross entropy as the cost function and Adam as the optimizer. In addition, for the model training process, we use a batch size of 512 and early stopping for 10 epoch patience.

## 2.3 Hidden Markov Model

We apply two types of supervised HMM as baseline models. First is a simple HMM which is a generative model and the other is hybrid deep neural network–HMM (DNN-HMM) which is a sequence-discriminative model [18].

### 2.3.1 Simple HMM

The simple HMM consists of three parameters: initial state distribution, transition probability and emission probability. In our case, the initial state distribution is the histogram of each chord in our train set. The transition probability is computed using the bigram of chord transition and it is assumed to follow the rule of general first-order Markov chains. A higher-order transition probability is not taken into account because the fixed length of an input bar in our task is not long enough. The emission probability is determined by a multinomial distribution of semitone observations from each chord class.

Once the parameters are learned, the model can generate a sequence of hidden chord states from a melody with three steps. First, the probabilities of 24 chord classes in each bar are determined by the melody distribution in each bar. As mentioned above, the simple HMM is a generative model. Hence, it uses not only the emission probability but also a class prior to calculate posterior probability with the Bayes rule. We define the class prior same as the initial probability, which is the histogram of each chord. Secondly, in order to reflect sequential effects, transition probability is applied to adjust the probabilities of the chord classes. In case of the first chord state, since there is no previous state to consider the transition, the initial probability is applied instead. After that, a Viterbi decoding algorithm is implemented to find the optimal chord sequence that is most likely to match along with the observed melody sequence [19].

### 2.3.2 DNN-HMM

The hybrid DNN-HMM is a popular model in the field of speech recognition [20]. It is a sequence-discriminative model, which adapts the advantage of sequential modeling method of HMM, but does not require the class prior and the emission probability to get posterior probability. DNN makes it possible because the probability result from a softmax output layer can be assumed as a posterior probability. Then the two of HMM parameters - initial state distribution and transition probability – are applied identically with the simple HMM to employ the Viterbi decoding algorithm.

We build an input layer with 12 units, 3 hidden layers with 128 units that are all identical and an output layer with 24 units. We use hyperbolic tangent activation function for the hidden layer and softmax for the output layer. Other features such as dropout, loss function, optimizer and batch size are applied in the same settings of BLSTM.

## 3. EXPERIMENTS

In this section, we first introduce our dataset, which is parsed from digital lead sheets. Then we present the experimental setup for evaluating the performance of chord generation models. We conduct both quantitative and qualitative evaluations for this task.

### 3.1 Dataset

We use the lead sheet database provided by Wikifonia.org, which was a public lead sheet repository. The site unfortunately stopped service in 2013, but some of the data, which consists of 5,533 Western music lead sheets in MusicXML format, including rock, pop, country, jazz, folk, R&B, children's song, etc., was obtained before the termination and we extracted features from the data for only academic purpose. From the obtained database, we collect 2,252 lead sheets, which are all in major key, and the majority of the bars in the lead sheets have a single chord per bar. If a bar consists of two or more chords, we choose the first chord in the bar. Then we extract musical features and convert them to a CSV format (see Section 2.1). The set is split into two sets – a training set of 1802 songs, which consists of 72,418 bars and a test set of 450 songs, which consists of 17,768 bars. Since musical features in this dataset can be useful for not only chord generation but also for other kinds of symbolic music tasks, the dataset is shared on our website (http://marg.snu.ac.kr/chord_generation/) for public access.

### 3.2 Quantitative Evaluation

We perform a quantitative analysis by comparing the accuracies of chord estimation from each model using the test set. The accuracy is calculated by counting the number of matching samples between the predicted and the true chords and by dividing it by the total number of samples. We mainly apply a 4-bar melody input for our task, but also experiment with 8-, 12- and 16-bar inputs to analyze the influence on the length of a melody sequence.

Determining the "right" chord is a difficult process because chord selection can vary among people based on their musical styles and tastes. However, the aforementioned accuracy calculation is often used to evaluate the capability of incorporating the long-term dependency in the musical progression [6, 8]. Therefore, we use it for measuring which model reflects the relationship between chord and melody most adequately.

### 3.3 Qualitative Evaluation

As mentioned above, there is a limit to evaluate the model performance only by a quantitative analysis. Thus, we also conduct qualitative evaluation based on subjective rating from actual user. This assessment allows us to determine the validity of each model by comparing how the chords generated from different models are perceived by actual users. For the experiment, we collect eighteen 4-bar-length melodies from lead sheets of thirteen K-pop songs and five Western pop songs. Every melodic sequence is converted into a vector of 12 semitones as described in Section 2.1. HMM, DNN-HMM, and BLSTM then generate chord sequences from each vector. Those sequences are evaluated by 25 musically untrained participants (13 males and 12 females) through a web-based survey.

The participants complete 18 sets of surveys in their own pace. At the beginning of each set, participants listen to a melody. After that, participants listen to the four types of chord progressions, including the one from the original song, along with the melody. Participants are asked to rate each chord progression on a five-point scale (1 – 'not appropriate'; 5 – 'very appropriate'). At the end of each set, participants also are asked to answer a question whether they have pre-existing familiarity with the original songs. The audio samples used for experiment are available on our website.

| # of Bars | HMM (%) | DNN-HMM (%) | BLSTM (%) |
|---|---|---|---|
| 4 | 40.33 | 45.02 | 50.55 |
| 8 | 40.43 | 44.82 | 50.32 |
| 12 | 40.41 | 44.95 | 49.23 |
| 16 | 40.45 | 44.68 | 49.90 |
| Average | 40.41 | 44.87 | 50.00 |

**Table 1**. Chord prediction performance using different number of input bar.

## 4. RESULTS

### 4.1 Chord Prediction Performance

Table 1 presents the accuracy results of three models for four instances of different bar lengths. The results show that the BLSTM method achieves the best performance on the test set followed by DNN-HMM and HMM. According to the average scores of models, BLSTM has 23.8% and 11.4% performance increase from the HMM and DNN-HMM, respectively. The results also demonstrate that the number of input bars is not an important factor affecting the accuracy for all models since they don't show obvious linear variations.

To examine the quality of predicted chords from each model more in depth, we compute the results of each model into a confusion matrix. This allows us to easily analyze the results through visualization. We normalize the matrix with the number of samples in respective chords so that each row represents the distribution of predicted chords on each true chord class. In Figure 3, we display this normalized confusion matrix of each model.

A number of noteworthy findings from each matrix are observed. First, HMM yields a skewed result that shows severe misclassification of chords especially on C, F and G as shown in Figure 3(a). We hypothesize this is resulted from the lack of complexity of the model. Emission probability, one of the parameters of the model, does not properly capture the accurate correlation between the chords and corresponding melodies. Moreover, the fact that the training data contains more frequent occurrences of C, F and G chords (over 60% in total samples) reduced the accuracy of the HMM model which uses the prior probability to obtain the posterior as mentioned in Section 2.3.1. Lastly, a noticeable bias in transition matrix moving to C chord also seems to lower the precision of the model.

The result of DNN-HMM is similar to HMM but the skewness on C chord spreads out little bit to F and G chords. Despite our initial expectation that the DNN would perform better since it is a discriminative model that calculates posterior directly, still many misclassifications on three chords exist as shown in Figure 3(b). To find the reasoning behind this observation, we test simple DNN with 1-bar input without the sequential parameter of HMM. The accuracy is higher than DNN-HMM (46.93%) and the confusion matrix produces more diagonal elements as shown in Figure 4. This finding supports that the transition probability of HMM forces the model to generate limited classes and also that the model is not adequate to train various chord progressions.

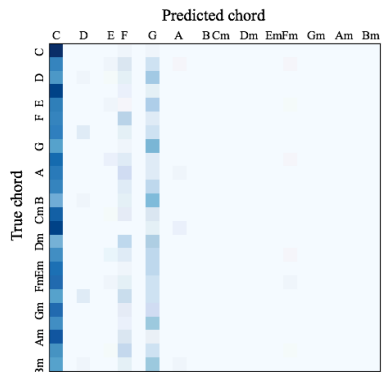

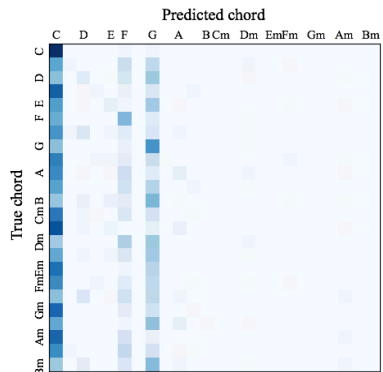

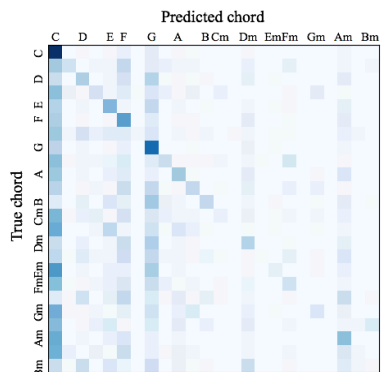

**Figure 3**. Normalized confusion matrix of HMM(a), DNN-HMM(b), and BLSTM(c) using 4-bar melody input.

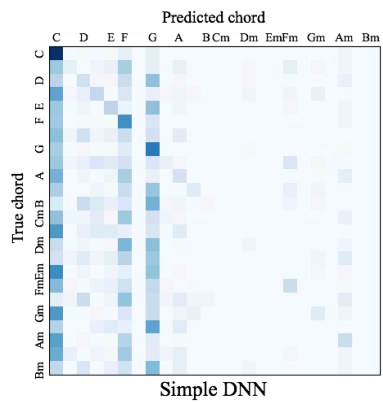

**Figure 4**. Normalized confusion matrix of simple DNN using single bar melody input.

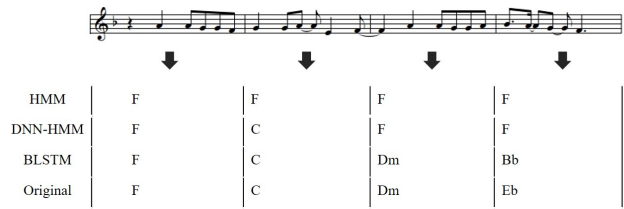

**Figure 5.** An example of generated chord progressions from three different models and the original progression.

In contrast to the HMM based method, the confusion matrix of the BLSTM shows a less skewed distribution and clearer diagonal elements as shown in Figure 3(c). BLSTM has much more complex parameters in hidden layers, which train the sequential information of both melodies and chords. We believe this property makes the performance better compared to the others.

### 4.2 User Preference

In the user subjective test, evaluation scores are obtained from 450 sets (18 sets x 25 participants). Each set contains chord sequences from HMM, DNN-HMM, and BLSTM. An original chord sequence is also included for relative comparison of the generated results to the original. These four chord sequences are evaluated as described in Section 3.3. Figure 5 shows the example of melody and chord sequences which is used in the user test and more examples are available to listen on our website.

The average score of each model is shown in Figure 6. The original chord progression is preferred the most followed by BLSTM, DNN-HMM, and HMM. To investigate whether differences on scores between the results are critical, we conduct one-way repeated measure ANOVA setting each model as a variable. The result shows that at least one out of four scores is significantly different from the others. ($F(3, 1772) = 310$, $p < 0.001$). We then conduct a pairwise t-test with Bonferroni correction on the mean scores between each pair of models for a post-hoc analysis. As a result, differences between all pairs are proven to be significant ($p < 0.01$). Therefore, it can be concluded that the BLSTM produces the most satisfying chord sequences among the other computational models but it produces less satisfying results than the original. Moreover, since the difference between BLSTM and DNN-HMM is bigger than other pairs, it seems there is a big quality difference between them.

To verify our hypothesis that having familiarity with the original song affects the result we perform a further analysis. We separate 450 evaluation sets into two, 248 sets marked as known and the rest as unknown, and conduct further analysis. A simple comparison of those two sets based on the evaluation scores shows that awareness of the songs does not affect the preference rank of the models. We also perform one-way repeated measure

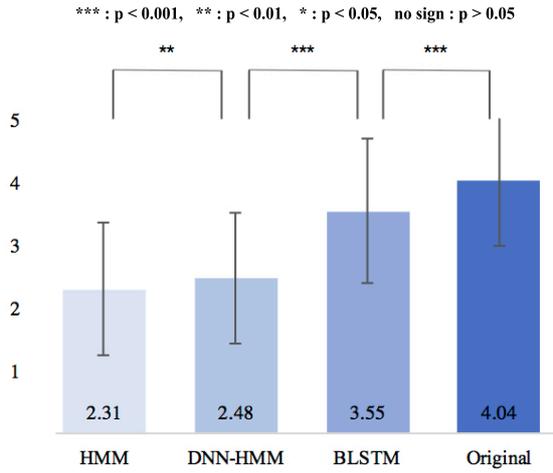

**Figure 6**. Mean score of subjective evaluation of each model.

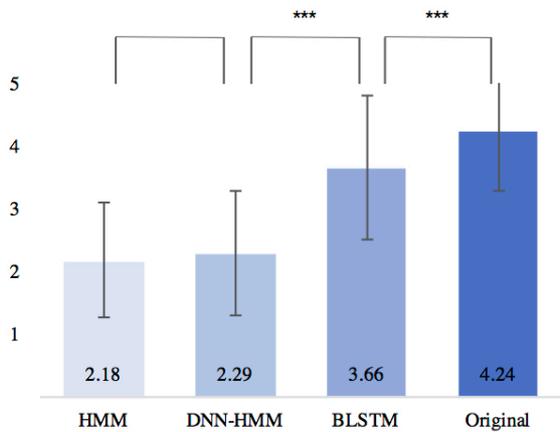

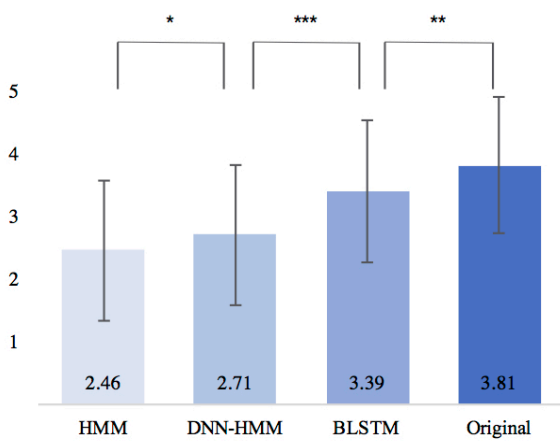

**Figure 7**. Mean score of subjective evaluation for a group of known songs (a), and of unknown songs (b).

ANOVA for each group of awareness (group of known songs: $F(3, 964) = 286, p < 0.001$ ; group of unknown songs: $F(3, 780) = 72, p < 0.001$) and pairwise t-test with Bonferroni correction. The results are presented in Figure 7. As shown in the figure, when songs are unknown, the preference for HMM based models increases while it decreases for BLSTM generated and original chords. A plausible explanation for this observation can be that when the listener knows the song, he/she is more perceptive of the monotonous chord sequences generated from HMM and DNN-HMM which tend to produce more of C, F and G than other chords. However, when the listener does not know the song, he/she is less aware of the monotonous progression of the chords and tend to give more generous scores to those two models. For BLSTM, the result is the opposite. Listeners who are more used to the dynamic chord progression of the original song tend to give relatively higher scores to BLSTM than to HMM based methods probably because BLSTM often generates a more diverse chord sequences. On the other hand, when the songs are unknown, relative preference towards both BLSTM and the original chords is less strong. The reduced gap among four different options when the songs are unknown may be explained by the assumption that when the songs are not familiar, all four options are relatively equally acceptable to the listeners. Regardless of the difference in the results, however, BLSTM is preferred over the other two models in both cases.

## 5. CONCLUSIONS

We have introduced a novel approach for generating a chord sequence from symbolic melody using neural network models. The result shows that BLSTM achieves the best performance followed by DNN-HMM and HMM. Therefore, the recurrent layer of BLSTM is more appropriate to model the relationship between melody and chord than HMM based sequential methods.

Our work can be further improved by modifying data extracting and preprocessing steps. First, since the lead sheets used in this study have one chord in each bar, the task is constrained to one-chord generation for each bar. Since actual music usually contains a lot of bars with multiple chords, additional extraction process is needed to allow the model to generate multiple chords per bar. Secondly, in the preprocessing step, all chords are mapped into only 24 classes of major and minor. Thus, further chord classes such as maj7 and min7 need to be included for performance improvement. Lastly, our input feature vectors consist of 12 semitones by accumulating the melody notes in each bar, so the sequential information of melodies in each bar disappears in this step. Thus, another feature-preprocessing step may be needed not to omit the information, which can be a crucial factor in the future work. We hope that more researches will be done through our published data to overcome the limitations as well as to further develop of this task.

## 6. ACKNOWLEDGEMENTS

This work was supported by Kakao Corp. and Kakao Brain Corp.